\documentclass[conference]{IEEEtran}
\IEEEoverridecommandlockouts
\usepackage{cite}
\usepackage{amsmath,amssymb,amsfonts}
\usepackage{algorithmic}
\usepackage{graphicx}
\usepackage{textcomp}
\usepackage{xcolor}
\def\BibTeX{{\rm B\kern-.05em{\sc i\kern-.025em b}\kern-.08em
    T\kern-.1667em\lower.7ex\hbox{E}\kern-.125emX}}

\usepackage{hyperref}
\usepackage{url}
\usepackage{amsmath}
\usepackage{xcolor}
\usepackage{amsfonts}
\usepackage{multirow}
\usepackage{verbatim}
\usepackage{tikz}
\usepackage{tikz,ifthen,pgfplots}
\usepackage{tcolorbox}
\usepackage[linesnumbered,ruled,vlined]{algorithm2e}
\usepackage{graphicx} 

\usetikzlibrary{arrows,trees,backgrounds,automata,shapes,decorations,plotmarks,fit,calc,positioning,shadows,chains}
\tikzstyle{every pin edge}=[<-,shorten <=1pt]
\tikzstyle{neuron}=[circle,fill=black!25,minimum size=17pt,inner sep=0pt]
\tikzstyle{input neuron}=[neuron, fill=green!50]
\tikzstyle{output neuron}=[neuron, fill=red!50]
\tikzstyle{hidden neuron}=[neuron, fill=blue!50]
\tikzstyle{annot} = [text width=4em, text centered]
\usetikzlibrary{calc}

\SetKwInput{KwInput}{Input}                
\SetKwInput{KwOutput}{Output}
\tikzstyle{line} = [draw, -latex']
\tikzstyle{decision} = [diamond, draw, 
text width=4.5em, text badly centered, node distance=3cm, inner sep=0pt]
\tikzstyle{block} = [rectangle, draw, 
text width=6em, text centered, rounded corners, minimum height=4em, ]
\tikzstyle{varblock} = [rectangle, draw, 
text width=6.5em, text centered, rounded corners, minimum height=4em]
\tikzstyle{highlightblock} = [rectangle, draw, fill=gray!45, 
text width=6.5em, text centered, rounded corners, minimum height=4em]
\tikzstyle{bigblock} = [rectangle, draw, fill=gray!20, 
text width=6.6em, text centered, rounded corners, minimum height=13.2em]
\tikzstyle{line} = [draw, -latex']
\tikzstyle{cloud} = [draw, ellipse,fill=red!20, node distance=3cm,
minimum height=2em]

\newcommand{\method}{\textit{DeepGalaxy }}

\begin{document}

\title{\textit{DeepGalaxy}: Testing Neural Network Verifiers via Two-Dimensional Input Space Exploration\\
}

\author{
	\IEEEauthorblockN{Xuan Xie}
	\IEEEauthorblockA{
	\textit{University of Alberta}\\
	Edmonton, Alberta\\
	lebron716@outlook.com}
	\and
	\IEEEauthorblockN{Fuyuan Zhang}
	fuyuanzhang@163.com
}

\maketitle

\begin{abstract}
Deep neural networks (DNNs) are widely developed and applied in many areas, and the quality assurance of DNNs is critical. 
Neural network verification (NNV) aims to provide formal guarantees to DNN models.
Similar to traditional software, neural network verifiers could also contain bugs, which would have a critical and serious impact, especially in safety-critical areas.
However, little work exists on validating neural network verifiers.
In this work, we propose \textit{DeepGalaxy}, an \textit{automated} approach based on \textit{differential testing} to tackle this problem.
Specifically, we (1) propose a line of mutation rules, including model level mutation and specification level mutation, to effectively explore the two-dimensional input space of neural network verifiers; and (2) propose heuristic strategies to select test cases.
We leveraged our implementation of \textit{DeepGalaxy} to test three state-of-the-art neural network verifies, Marabou, Eran, and Neurify. 
The experimental results support the efficiency and effectiveness of \textit{DeepGalaxy}.
Moreover, five unique unknown bugs were discovered.
\end{abstract}


\section{Introduction}
\label{Sec:intro}

Deep neural networks are becoming more and more popular due to their remarkable performance in dealing with challenging problems, such as machine translation \cite{machinetranslation}, speech recognition \cite{speechrecognition}, and autonomous driving \cite{autonomousdriving1, autonomousdriving2}.
%Despite their tremendous success, neural networks are known to be vulnerable to adversarial examples, which are carefully designed inputs aiming to fool neural networks into making wrong predictions \cite{advexample1, advexample2}. 
Despite their tremendous success, it is difficult to formally verify that neural networks satisfy desired specifications. This hinders the deployment of neural networks in safety-critical domains. 

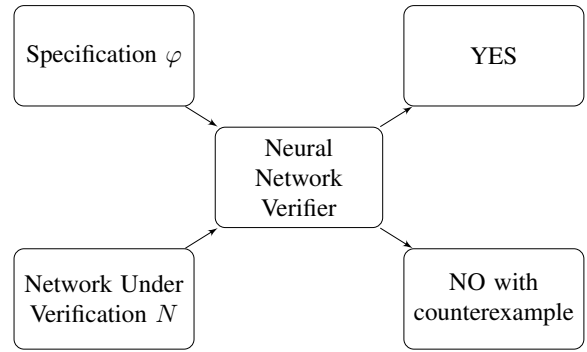
\begin{figure}[t]
	\centering 
	\scalebox{.95}{
		\begin{tikzpicture}[node distance = 1.6cm, auto]
			\node [block] (verification) {Neural Network Verifier};
			\node [varblock, below left = 0.4cm of verification] (nuv) {Network Under Verification $N$};
			\node [varblock, above left = 0.4cm of verification] (spec) {Specification $\varphi$};
			\node [varblock, above right = 0.4cm of verification] (proof) {YES};
			\node [varblock, below right = 0.4cm of verification] (ce) {NO with counterexample};
			
			\path [line] (nuv) -- (verification);
			\path [line] (verification) -- (proof);
			\path [line] (verification) -- (ce);
			\path [line] (spec) -- (verification);
		\end{tikzpicture}
	}
	\caption{Overview of Neural Network Verification}
	\label{fig:nnv}
\end{figure}

In order to mitigate this problem, neural network verification (NNV) \cite{reluplex, ai2} is proposed to %make sure DNNs run correctly and giving formal guarantees to neural network models. 
verify properties of neural networks, e.g., adversarial robustness \cite{advexample1, advexample2}, and aims at providing formal guarantees to DNNs.
Figure \ref{fig:nnv} shows the overview of NNV.
Given a network $N$ under verification and a specification $\varphi$,
neural network verification seeks an answer for the question: whether the network $N$ satisfies $\varphi$?
%Given the aforementioned input, the neural network verifier tries to return a counterexample that violates the desired specification.
%Otherwise, a proof of correctness of $\varphi$ is returned.
Given the aforementioned input, neural network verifiers can either provide a proof of correctness of $\varphi$, or find a counterexample that violates the specification $\varphi$.
Recently, there has been a plethora of research on neural network verification.
For instance, adversarial robustness certification \cite{reluplex, ai2} tries to prove the absence of adversarial examples, and fairness analysis \cite{libra} verifies whether DNN models satisfy certain fairness properties.

Nevertheless, like all traditional software, NNV implementations (i.e., neural network verifiers) also contain bugs and may return wrong verification results.
%As neural network verifiers are used to give formal guarantees to the DNN models, 
%Bugs in neural network verifiers may lead to fatal errors for DNNs used in safety-critical domains, e.g., autonomous driving.
It is possible that fatal errors of DNNs still remain after verification, which could cause sever consequences for DNNs in safety-critical domains, e.g., autonomous driving. However, no previous work exists on automatically and systematically giving quality assurance of neural network verifiers.

In this work, we aim to validate the correctness of neural network verifiers.
Nevertheless, there are several challenges, and we see the main challenges for testing neural network verifiers as being two-fold.
(1) Neural network verifier testing remains largely untouched, and challenges come while generating test suite.
Different from classical software testing, which takes files or concrete values as input and is typically one dimensional, the inputs to neural network verifiers are  DNN models and desired specifications, as shown in Figure \ref{fig:nnv}. 
Thus, the input space to the verifiers has two dimensions, which makes the input space exploration more difficult.
There is a lack of systematic methods to generate effective test cases to explore verifiers' behavior.
(2) There is no oracle to distinguish whether indeed neural network verifiers give wrong verification results.
Although in traditional software testing, there are many test oracles to identify if a real bug is detected, it is hard to expose the bug in neural network verifier testing because of the absence of test oracles.

Towards solving the aforementioned challenges, we propose \textit{DeepGalaxy}, a novel approach for detecting bugs in neural network verifiers.
\textit{DeepGalaxy} is designed to overcome the above mentioned challenges.
To address the first challenge, we design two levels of mutation operators to produce various test cases to explore the two-dimensional input space.
In particular, we propose (1) DNN model level mutation, which mutates the weights and activation functions in each layer of the model, (2) specification level mutation, which mutates the specification of interest.
With the help of two-dimensional mutations, \method is able to cover more behavior of the verifiers under test, and expose the underlying bugs.
To address the second challenge, we leverage the classical idea of \textit{differential testing} to identify potential bugs.
In other words, we input the same DNN models and specifications to multiple neural network verifiers, and observe whether there are inconsistencies in the output of the verifiers. If one verifier gives an answer different from all the other verifiers, we consider that as a bug.

To sum up, this paper makes the following key contributions:
\begin{itemize}
	\item We propose a novel blackbox mutational testing technique, \textit{DeepGalaxy}, for detecting bugs in neural network verifiers. To the best of our knowledge, our work is the first to automatically validate the correctness of neural network verifiers.
	\item We design a set of mutation rules to effectively and efficiently generate DNN model mutants and specification mutants, which facilitates the test case generating process for triggering bugs in neural network verifiers.
	\item We implemented \textit{DeepGalaxy}, tested three state-of-the-art neural network verifiers (Marabou, ERAN, Neurify), and found five unique bugs.
\end{itemize}

\section{Background and Problem Definition}
\label{Sec:problemdef}

In this section, we introduce the background and define the problems that we aim to solve.

\noindent \textbf{Deep Neural Networks.}
A deep neural network (DNN) consists of multiple layers, including an input layer, one or more hidden layers, and an output layer. 
Each of the layers is composed of multiple neurons, and each neuron gets input values from the previous layer and gives the computed output value to the next layer.
Typically, the computation of a neuron includes an affine transformation followed by a non-linear transformation, which is called activation function.
Some popular activation functions are ReLU and Sigmoid.
There are many kinds of DNN models, e.g., Convolutional Neural Networks (CNNs), which is good at image classification, and Recurrent Neural Networks (RNNs), which performs well in audio recognition.

\noindent \textbf{Neural Network Verification (NNV).} NNV \cite{reluplex, ai2} aims at making sure DNNs run correctly and giving formal guarantees to neural network models.
For a DNN model $N$ and a specification $\phi$, a neural network verifier checks whether $\phi$ is satisfied on $N$ for all inputs $x$. The specification $\phi$ holds if there are no counterexamples. 
Otherwise, the specification is violated and a counterexample is returned by the verifier. 

A specification can be expressed in the form: $$\forall x, \phi_{pre}(x) \land y \leftarrow N(x) \implies \phi_{post}(x, y).$$ 
Here, we use $\phi_{pre}(x)$ to specify the preconditions on the input of $N$. We use $\phi_{post}(x, y)$ to express the postconditions on the input and the output of $N$. We use $y \leftarrow N(x)$ to encode the process of calculating the output of $N$ for input $x$ and assigning the output value to $y$.

% Given a DNN model $N$ and a specification $\varphi$, a neural network verifier verifies whether $N$ satisfies $\varphi$ on all inputs $x$. 
% The specification holds if no counter-example exists. 
% Otherwise, a counter-example, found by the neural network verifier, is returned.
% Typically, $\varphi$ is of the form:
% \begin{align*}
% 	\{\varphi_{pre}(x)\}\\
% 	y \leftarrow N(x)\\
% 	\{\varphi_{post}(x, y)\},
% \end{align*}
% where $\{\varphi_{pre}(x)\}$ is the preconditions over the input of $N$, $\{\varphi_{post}(x, y)\}$ is the postconditions over the input and the output of $N$, and $y \leftarrow N(x)$ refers to the procedure of computing the output of $N$ for input $x$ and assigning the output value to $y$.
% The meaning of the above triple is equivalent to the implication: $\forall x, \varphi_{pre}(x) \land y \leftarrow N(x) \implies \varphi_{post}(x, y)$.
%This area is quite new, but there are many popular tools with different techniques. 
Recent researches have already developed many verification techniques. 
For example, Planet \cite{planet} encodes the linear approximation of the overall network behavior into SMT or integer linear programming problems, and
Zhang et al. \cite{crown} proposed CROWN, which computes the certified lower bound of minimum perturbation in DNNs for any given data point.

\noindent \textbf{Problem Definition.}
Similar to the traditional software, neural network verifiers could also contain bugs, which is fatal in safety-critical domains.
%However, different from traditional software testing, the inputs to neural network verifiers contain two parts, the formal specification, which is the properties to be verified, and DNN models.
Notice that there are two inputs to neural network verifiers, the formal specification, i.e., the properties to be verified, and DNN models.
We define the \textit{test cases} in neural network verifier testing as follows.

\vspace{0.1cm}
\textbf{Definition 1 (Test Case).} \textit{A test case for neural network verifier testing is a tuple ($N$, $\varphi$), where N is a DNN model and $\varphi$ stands for the desired specification.}
\vspace{0.1cm}

Having defined the test cases in neural network verifier testing, we now describe the main problem we target at.

\vspace{0.1cm}
\textbf{Problem 1 (Neural Network Verifier Testing).} \textit{Given a neural network verifier V, generate a set of valid test cases T to detect bugs in V.}
\vspace{0.1cm}

In general, this problem is very challenging because of the intractability of the input space exploration and the absence of the test oracle. 
In the following section, we propose a novel technique, called \textit{DeepGalaxy}, to tackle this problem. 

\section{Approach}
\label{Sec:mutation}
In this section, we describe our testing framework, \textit{DeepGalaxy}, in more details.

\begin{figure}[t]
	\centering 
	\scalebox{2.8} {\includegraphics[width=3cm]{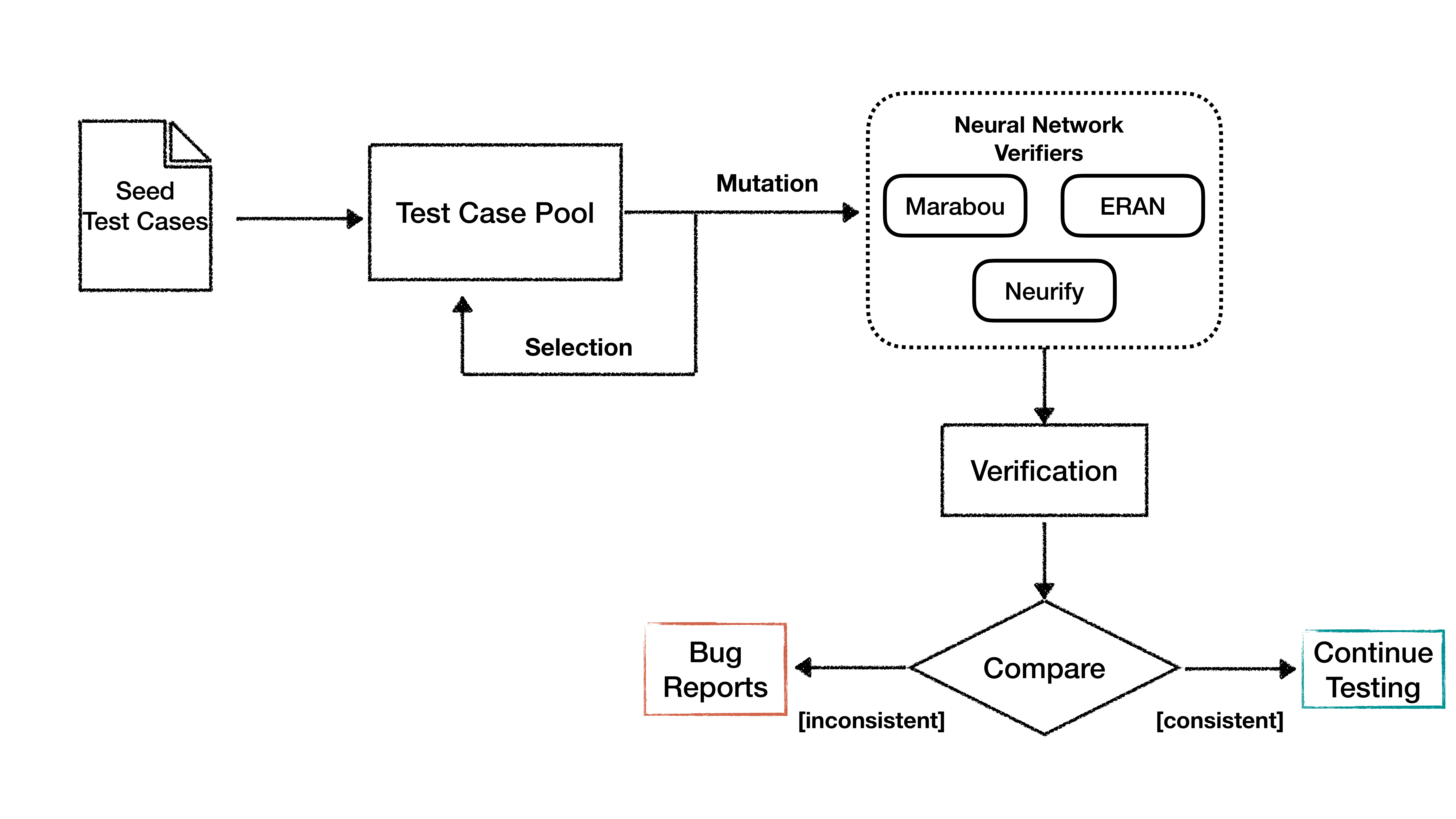}}
	\caption{Overview of our testing framework \textit{DeepGalaxy}.}
	\label{fig:framework}
\end{figure}
 
Figure \ref{fig:framework} presents the overall workflow.
Given a set of seed test cases $T_{seed}$, \method produces potential bugs in the neural network verifiers. 
First, we put $T_{seed}$ into the test case pool \textit{P}.
Then, the test cases in \textit{P} are \textbf{selected} based on designed heuristics.
The test case $(N, \varphi)$ is \textbf{mutated} by using different levels of mutation operators, and neural network verifiers verify whether $\varphi$ holds for \textit{N}.
In the end, the verification results from different verifiers under test are compared with each other. If one of the results from verifier \textit{V} is inconsistent than the others, a \textbf{bug} is found.
In the following three subsections, we describe the predominant parts of \textit{DeepGalaxy}: \textit{test case selection}, \textit{mutation}, and \textit{bug report.}

\subsection{Test Case Selection}
\label{subsec:select}
Since the two-dimensional input space, including DNN model space and specification space, is prohibitively large, even infinite.
It is infeasible to explore the space completely.
Therefore, we design three model selection approaches to generate the DNN models and the specification towards the direction of exposing the potential bugs in various neural network verifiers.
\begin{itemize}
	\item \textit{Random Selection:} to randomly select a test case from the test case pool, which utilizes the advantage of exploration of the space.
	\vspace{0.05cm}
	\item \textit{Recency-aware:} to select the test case that is last added to the pool.
	\vspace{0.05cm}
	\item \textit{Mixed:} to combine the recency-aware selection and the random selection, which balance exploration and exploitation. We randomly choose recency-aware selection and random selection.
\end{itemize}
With the help of the model selection approach, the test case that can exploit the unused code in the verifier under test can be selected more frequently.
Next, we introduce the two levels of mutation, DNN model level mutation and specification level mutation, which is the key to expose the underlying bug.

\subsection{Mutation}
\label{subsec:mutation}
The input space to be explored is extremely large, since that includes the space of DNN models and the specifications.
In order to explore the space thoroughly, we propose two levels of mutation operators.

\noindent \textbf{DNN Model Level Mutation.}
The goal of the model level mutation is to test neural network verifiers by exploring the underlying un-triggered operation of the verifier or diverse combination of behaviors of the verifier.
In particular, we adopt five mutation rules from existing work \cite{deepmutation} to fulfill our mutation task.

\begin{itemize}
	\item \textit{Gaussian Fuzzing (GF):} to fuzz the value of the weights following the Gaussian distribution $N(\mu, \delta^2)$, where $\mu$ is the mean value and $\delta$ is the standard deviation. We set $\mu$ as zero and $\delta$ as 0.5.
	\vspace{0.05cm}
	\item \textit{Weight Shuffling (WS):} to randomly choose a neuron and shuffles the weights of its connections to the previous layer.
	\vspace{0.05cm}
	\item \textit{Neuron Effect Blocking (NEB):}  to block neuron effects to all of the connected neurons in the next layers by resetting its connection weights of the next layer to be zero.
	\vspace{0.05cm}
	\item \textit{Neuron Activation Inverse (NAI):} to change the sign of the output value before giving it to the corresponding activation function. The goal is to invert the activation status of a neuron.
	\vspace{0.05cm}
	\item \textit{Neuron Switch (NS):} to randomly switch two neurons within a layer	to change their effects to the next layers.
\vspace{0.05cm}
\end{itemize}

\vspace{0.1cm}

\noindent \textbf{Specification Level Mutation.}
We leverage four newly designed mutation rules for specification level mutation.

\begin{itemize}
	\item \textit{Constant Addition (CA):} to select a constant in the specification and add a random value to it.
	\vspace{0.05cm}
	\item \textit{Constant Removal (CR):} to remove a constant in the specification.
	\vspace{0.05cm}
	\item \textit{Constant Subtraction (CSb):} to select a constant in the specification and subtract a random value on it.
	\vspace{0.05cm}
	\item \textit{Constant Switch (CSw):} to switch two constants in the specification.
\end{itemize}

We take adversarial robustness as an example to illustrate how to conduct specification level mutation.
Suppose that we have a neural network $N$ trained on MNIST dataset and we want to verify adversarial robustness on $N$.
Given the DNN $N$, a fixed input $x'$, a region $R \subset \mathbb{R}^n$, distance function $dist$, and $\varepsilon = 0.1$, the seed adversarial robustness specification is
\begin{equation}
	\label{eq1}
	\forall x \in R: dist(x, x') \leq 0.1 \implies N(x) = N(x').
\end{equation}
To mutate the above specification using the mutation rule of CA, we randomly change the value of $\varepsilon$ to 0.15.
The mutated specification becomes
\begin{equation}
	\forall x \in R: dist(x, x') \leq 0.15 \implies N(x) = N(x').
\end{equation}
This new mutated specification can then be used as the input to the verifiers under test.

\subsection{Bug Identification}
\label{subsec:bugiden}
One of the main challenges of testing neural network verifiers is that it is difficult to expose and identify the potential bugs.
Different from classical software testing, where typically there are oracles to indicate the ground truth, there exist no oracles to distinguish whether the detected bug is a real bug.
Therefore, we leverage differential testing to identify whether a real bug is detected.
In particular, \method compares the results returned from different verifiers. If one of them is inconsistent with the others, we consider that a bug is detected.
Besides the inconsistency bugs, we also consider crash, e.g., the neural network verifier exits abnormally, as a kind of bug.

\subsection{Algorithm}
Algorithm \ref{algo} shows our neural network verifier testing approach to capturing underlying bugs.
The inputs are a seed DNN model $N$, a seed specification $\varphi$, the set of verifiers under test $V$, maximum iteration \textit{maxIter}, and test case selection method \textit{selection}, as described in subsection \ref{subsec:select}.
The output is the bug report $b$.

First, we initialize the number of iteration \textit{iter}, the result set $R$, and the test case pool $P$ (Line 1-3).
Then, \method iterates for \textit{maxIter} times (Line 4).
In each iteration, it selects the test case from $P$ (Line 6), and performs the mutation by randomly selecting the mutation operator (Line 7).
For each test case, \method input it to each verifier and collect the result (Line 8-11).
If any of the results is inconsistent with the other results, report the bug (Line 12-14).
Otherwise, add $T$ to the test case pool.
 
\begin{algorithm}
	\label{algo}
	\caption{Neural Network Verifier Testing}
	\KwInput{$N$: seed DNN model, $\varphi$: seed specification, $V$: neural network verifiers under test, \textit{maxIter}: maximum iterations, \textit{selection}: test case selection method}
	\KwOutput{$b$: bug report}
	
	$iter = 0$;\\
	$R = \{\};$\\
	$P  = \{(N, \varphi)\}$;\\
	\While{$iter < maxIter$}{
		$iter = iter + 1;$\\
		$t = selection(P)$;\\
		$T = mutation(t)$;\\
		\For{$t' \in T$ }{
			\For{$v \in V$}{
				$result = v(t');$\\
				$R = R $ $\cup$ $\{result\};$
			}
			\For{$r \in R, r' \in R$}{
				\If{$r \neq r'$}{
					\Return b;
				}
			}
		$R = \{\};$\\
		}
		$P = P$ $\cup$ $T;$
	}
\end{algorithm}

\section{Evalutation}
\label{Sec:evaluation}

In this section, we describe the experimental evaluation of \textit{DeepGalaxy}. In particular, we investigate the following research questions:

\begin{itemize}
	\item \textbf{RQ1:} How efficient is \method in generating test cases?
	\item \textbf{RQ2:} Can \method  detect real bugs in neural network verifiers?
\end{itemize}

\noindent \textbf{Experimental Setup.} 
We select three state-of-the-art neural network verifiers as the bug-hunting object.
In particular, we choose (1) SMT-based tool Marabou (version \texttt{9b1dd78}) \cite{marabou}, which is an improved version of Reluplex \cite{reluplex} and it combines parallel processing and bound tightening to improve the efficiency of SMT solving,
(2) ERAN (version \texttt{ebeaa70}) \cite{eran}, which certifies adversarial robustness by transforming DNN analysis into the classical framework of abstract interpretation and uses multiple abstract domains, e.g., zonotope and polyhedra, for overapproximation,
(3) Neurify (version \texttt{31e9c3b}) \cite{neurify}, which integrates the idea of linear relaxation and abstraction refinement in interval propagation.

For the seed DNN models, we trained a fully connected DNN, with three hidden layers and 50 neurons per layer, on MNIST \cite{mnist} dataset.
As future work, we also plan to use popular DNN models, e.g., Alexnet\cite{alexnet}, as initial seed models.
For the seed specification, we choose adversarial robustness \cite{advrobustness} to conduct mutation, because it is a popular safety-critical specification that people pay attention to nowadays.
In particular, we use equation \ref{eq1} as the seed specification.

All experiments are conducted on a machine with Intel (R) Core (TM) i5 CPU @2.4 GHz and 16 GB RAM, equipped with a GNU/Linux system.

\vspace{0.1cm}
\noindent \textbf{RQ1: Efficiency of \textit{DeepGalaxy}.}

We investigate the efficiency of \textit{DeepGalaxy}, and use the execution time as the metrics.
On average, the execution time of each testing round of \method is 39.6 seconds.
Notice that some verification procedure takes more time, because the speed of different verifiers for tackling verification is different.
For example, ERAN, based on abstract interpretation, conducts fast but not complete verification, whereas Marabou, based on SMT solving, is relatively slow.

\vspace{0.1cm}
\begin{tcolorbox}[size=title]
	{ \textbf{Answer to RQ1:} \method is efficient in terms of generating mutation test cases, which indicates the usefulness when applying it in practice.} 
\end{tcolorbox}

\vspace{0.1cm}
\noindent \textbf{RQ2: Bug Hunting of \textit{DeepGalaxy}.}

We use bug hunting numbers as the measuring metrics in this research question.
In total, \method discovered five bugs in three state-of-the-art neural network verifiers, where two were confirmed.
More specifically, we captured two bugs in Marabou, two bugs in ERAN, and one bug in Neurify.
Table \ref{table:bugs} is overview of bugs found by \textit{DeepGalaxy}.
The numbers in the table represent the bugs that \method detected, confirmed, unconfirmed, and fixed in the three verifiers, respectively.
\textit{Detected} means the bugs found by \textit{DeepGalaxy}.
\textit{Confirmed} means the bugs have been confirmed by the developers.
\textit{Unconfirmed} means we reported the bugs to the developers, but have not received a confirmation yet.
\textit{Fixed} means the developers confirmed the bugs and fixed them.
After \method detects the bugs, we also use the manual inspection to facilitate bug reporting and enhance the quality of the bug report.

\vspace{0.1cm}
\begin{tcolorbox}[size=title]
	{ \textbf{Answer to RQ2:} Our approach can detect real bugs in the state-of-the-art neural network verifiers. Several bugs have been reported, and one bug has been fixed by the developer.} 
\end{tcolorbox}

\begin{table}[htbp]
	\caption{Bugs overview.}
	\begin{center}
		\begin{tabular}{|c|c|c|c|c|}
			\hline
			\textbf{Bugs}&\multicolumn{4}{|c|}{\textbf{Neural Network Verifiers}} \\
			\cline{2-5} 
			\textbf{Status} & Marabou& ERAN& Neurify & Total \\
			\hline
			Detected & 2& 2 & 1 & 5 \\
			\hline
			Confirmed & 0 & 2 & 0 & 2 \\
			\hline
			Unconfirmed & 2 & 0 & 1 & 3 \\
			\hline
			Fixed & 0 & 1 & 0 & 1 \\
			\hline
		\end{tabular}
		\label{table:bugs}
	\end{center}
\end{table}

\section{Related Work}
\label{Sec:relatedwork}

\noindent \textbf{Testing Classical Software Verifiers.}
In classical software, software verifiers are broadly utilized to enhance the quality assurance of software systems.
There is a considerable amount of study on the correctness and reliability of software verifiers.
Zhang et al. propose MCFuzz \cite{mcfuzz}, which is an automated fuzzing technique to test software model checkers, e.g., CPAChecker \cite{cpachecker}. 
It leverages branch reachability to tackle scalability and the oracle problem.
Yinyang \cite{yinyang} is designed to validate the correctness of SMT solvers, e.g., Z3 \cite{z3} and CVC4 \cite{cvc4}, via fusing pairs of equisatisfiable formulas.
Please note that, different from them, our work takes the first step to validate the correctness of neural network verifiers with both DNN model level mutation and specification level mutation.

\noindent \textbf{Testing Deep Neural Networks.} 
There are also many works on giving quality assurance on DNN models. 
For instance, Pei et al. \cite{deepxplore} propose to leverage neuron coverage to guide the generation of test cases.
Ma et al. \cite{deepgauge} propose multiple granularity coverage criteria to evaluate the adequacy of testing DNN models.
Odena et al. \cite{tensorfuzz} first design a coverage-guided fuzzing framework for testing deep neural networks. Zhang et al. \cite{deepsearch} develop a fuzzing-based blackbox adversarial attack for DNNs.
Moreover, Khmelnitsky et al. \cite{pdv} design an technique to extract a surrogate automata model to analyze and verify regular properties for recurrent neural networks.
Notice that our work is orthogonal to these methods, because we primarily focus on detecting bugs in neural network verifiers, whereas they focus on measuring the quality of DNN models.

\section{Conclusion}
\label{Sec:conclusion}

In this paper, we present a novel approach, \textit{DeepGalaxy}, to automatically and systematically validating neural network verifiers.
In particular, we design a set of mutation operators to generate effective DNN model mutants and specification mutants, so as to explore the two-dimensional input space.
Moreover, we leverage the classical idea of differential testing to capture the bugs in the neural network verifiers.
In the experiment, we demonstrate that \method is efficient in the test case generation and effective in bug hunting.

As future work, we plan to design more levels of mutation rules, e.g., configuration level, to explore more unused functionalities of neural network verifiers.
In addition, we plan to extend our scope of neural network verifiers selection and select more types of verifiers, e.g., mixed integer linear programming based verifiers, to perform a larger scale experiment.

%
%\begin{figure}[htbp]
%\centerline{\includegraphics{fig1.png}}
%\caption{Example of a figure caption.}
%\label{fig}
%\end{figure}

%\section*{Acknowledgment}

%\section*{References}


\begin{thebibliography}{00}
\bibitem{reluplex} Katz, G., Barrett, C., Dill, D. L., Julian, K., Kochenderfer, M. J. (2017, July). Reluplex: An efficient SMT solver for verifying deep neural networks. In International Conference on Computer Aided Verification (pp. 97-117). Springer, Cham.
\bibitem{ai2} Gehr, T., Mirman, M., Drachsler-Cohen, D., Tsankov, P., Chaudhuri, S., Vechev, M. (2018, May). Ai2: Safety and robustness certification of neural networks with abstract interpretation. In 2018 IEEE Symposium on Security and Privacy (SP) (pp. 3-18). IEEE.
\bibitem{marabou} Katz, G., Huang, D. A., Ibeling, D., Julian, K., Lazarus, C., Lim, R., Barrett, C. (2019, July). The marabou framework for verification and analysis of deep neural networks. In International Conference on Computer Aided Verification (pp. 443-452). Springer, Cham.
\bibitem{neurify} Wang, S., Pei, K., Whitehouse, J., Yang, J., Jana, S. (2018). Efficient formal safety analysis of neural networks. arXiv preprint arXiv:1809.08098.
\bibitem{eran} ERAN: https://github.com/eth-sri/eran
\bibitem{mcfuzz} Zhang, C., Su, T., Yan, Y., Zhang, F., Pu, G., Su, Z. (2019, August). Finding and understanding bugs in software model checkers. In Proceedings of the 2019 27th ACM Joint Meeting on European Software Engineering Conference and Symposium on the Foundations of Software Engineering (pp. 763-773).
\bibitem{yinyang} Winterer, D., Zhang, C., Su, Z. (2020, June). Validating SMT solvers via semantic fusion. In Proceedings of the 41st ACM SIGPLAN Conference on Programming Language Design and Implementation (pp. 718-730).
\bibitem{deepgauge} Ma, L., Juefei-Xu, F., Zhang, F., Sun, J., Xue, M., Li, B., Wang, Y. (2018, September). Deepgauge: Multi-granularity testing criteria for deep learning systems. In Proceedings of the 33rd ACM/IEEE International Conference on Automated Software Engineering (pp. 120-131).
\bibitem{pdv} Khmelnitsky, I., Neider, D., Roy, R., Xie, X., Barbot, B., Bollig, B., ... Ye, L. (2021, October). Property-Directed Verification and Robustness Certification of Recurrent Neural Networks. In 19th International Symposium on Automated Technology for Verification and Analysis (ATVA 2021).
\bibitem{z3} De Moura, L., Bjørner, N. (2008, March). Z3: An efficient SMT solver. In International conference on Tools and Algorithms for the Construction and Analysis of Systems (pp. 337-340). Springer, Berlin, Heidelberg.
\bibitem{cpachecker} Beyer, D., Keremoglu, M. E. (2011, July). CPAchecker: A tool for configurable software verification. In International Conference on Computer Aided Verification (pp. 184-190). Springer, Berlin, Heidelberg.
\bibitem{mnist} LeCun, Y., Cortes, C. (2010). MNIST handwritten digit database.
\bibitem{alexnet} Krizhevsky, A., Sutskever, I., Hinton, G. E. (2012). Imagenet classification with deep convolutional neural networks. Advances in neural information processing systems, 25, 1097-1105.
\bibitem{deepmutation} Ma, L., Zhang, F., Sun, J., Xue, M., Li, B., Juefei-Xu, F., Wang, Y. (2018, October). Deepmutation: Mutation testing of deep learning systems. In 2018 IEEE 29th International Symposium on Software Reliability Engineering (ISSRE) (pp. 100-111). IEEE.
\bibitem{advrobustness} Goodfellow, I. J., Shlens, J., Szegedy, C. (2014). Explaining and harnessing adversarial examples. arXiv preprint arXiv:1412.6572.
\bibitem{machinetranslation} Cho, K., Van Merriënboer, B., Gulcehre, C., Bahdanau, D., Bougares, F., Schwenk, H., Bengio, Y. (2014). Learning phrase representations using RNN encoder-decoder for statistical machine translation. arXiv preprint arXiv:1406.1078.
\bibitem{speechrecognition} Graves, A., Mohamed, A. R., Hinton, G. (2013, May). Speech recognition with deep recurrent neural networks. In 2013 IEEE international conference on acoustics, speech and signal processing (pp. 6645-6649). Ieee.
\bibitem{autonomousdriving1} Bojarski, M., Del Testa, D., Dworakowski, D., Firner, B., Flepp, B., Goyal, P., Zieba, K. (2016). End to end learning for self-driving cars. arXiv preprint arXiv:1604.07316.
\bibitem{autonomousdriving2} Xu, H., Gao, Y., Yu, F., Darrell, T. (2017). End-to-end learning of driving models from large-scale video datasets. In Proceedings of the IEEE conference on computer vision and pattern recognition (pp. 2174-2182).
\bibitem{advexample1} Szegedy, C., Zaremba, W., Sutskever, I., Bruna, J., Erhan, D., Goodfellow, I., Fergus, R. (2013). Intriguing properties of neural networks. arXiv preprint arXiv:1312.6199.
\bibitem{advexample2} Szegedy, C., Zaremba, W., Sutskever, I., Bruna, J., Erhan, D., Goodfellow, I.,  Fergus, R. (2013). Intriguing properties of neural networks. arXiv preprint arXiv:1312.6199.
\bibitem{libra} Urban, C., Christakis, M., Wüstholz, V., Zhang, F. (2020). Perfectly parallel fairness certification of neural networks. Proceedings of the ACM on Programming Languages, 4(OOPSLA), 1-30.
\bibitem{deepxplore} Pei, K., Cao, Y., Yang, J., Jana, S. (2017, October). Deepxplore: Automated whitebox testing of deep learning systems. In proceedings of the 26th Symposium on Operating Systems Principles (pp. 1-18).
\bibitem{tensorfuzz} Odena, A., Olsson, C., Andersen, D., Goodfellow, I. (2019, May). Tensorfuzz: Debugging neural networks with coverage-guided fuzzing. In International Conference on Machine Learning (pp. 4901-4911). PMLR.
\bibitem{crown} Zhang, H., Weng, T. W., Chen, P. Y., Hsieh, C. J., Daniel, L. (2018). Efficient neural network robustness certification with general activation functions. arXiv preprint arXiv:1811.00866.
\bibitem{planet} Ehlers, R. (2017, October). Formal verification of piece-wise linear feed-forward neural networks. In International Symposium on Automated Technology for Verification and Analysis (pp. 269-286). Springer, Cham.
\bibitem{cvc4} Barrett, C., Conway, C. L., Deters, M., Hadarean, L., Jovanović, D., King, T., Tinelli, C. (2011, July). Cvc4. In International Conference on Computer Aided Verification (pp. 171-177). Springer, Berlin, Heidelberg.
\bibitem{deepsearch} Zhang, F., Chowdhury, S., Christakis, M. (2020).
DeepSearch: a simple and effective blackbox attack for deep neural networks. In proceedings of the 28th ACM Joint European Software Engineering Conference and Symposium on the Foundations of Software Engineering. (pp.800-812). California, United States.
\end{thebibliography}
\end{document}